\documentclass[10pt,twocolumn,letterpaper]{article}

\usepackage{iccv}
\usepackage{times}
\usepackage{epsfig}
\usepackage{graphicx}
\usepackage{amsmath}
\usepackage{amssymb}
\usepackage[symbol]{footmisc}
\usepackage{multirow}

\usepackage[breaklinks=true,bookmarks=false]{hyperref}

\usepackage{capt-of,etoolbox}

\usepackage{pifont}
\newcommand{\cmark}{\ding{51}}%
\newcommand{\xmark}{\ding{55}}%

\newcommand{\HeadEvolver}{Our framework} 

\iccvfinalcopy 

\def\iccvPaperID{****} 

\ificcvfinal\fi

\makeatletter
\apptocmd\@maketitle{{\myfigure{}\par}}{}{}
\makeatother

\begin{document}

\title{HeadEvolver: Text to Head Avatars via Expressive \\ and Attribute-Preserving Mesh Deformation}

\author{Duotun Wang$^{1*}$, Hengyu Meng$^{1*}$, Zeyu Cai$^{1}$\thanks{indicates equal contributions}, Zhijing Shao$^{1}$, Qianxi Liu$^{1}$, Lin Wang$^{1, 3}$\\
Mingming Fan$^{1, 3}$, Xiaohang Zhan$^{2}$, Zeyu Wang$^{1, 3}$\\
$^{1}$The Hong Kong University of Science and Technology (Guangzhou) $^{2}$Tencent AI Lab\\  $^{3}$The Hong Kong University of Science and Technology\\
}

\newcommand\myfigure{%
\centering
  \includegraphics[width=\textwidth]{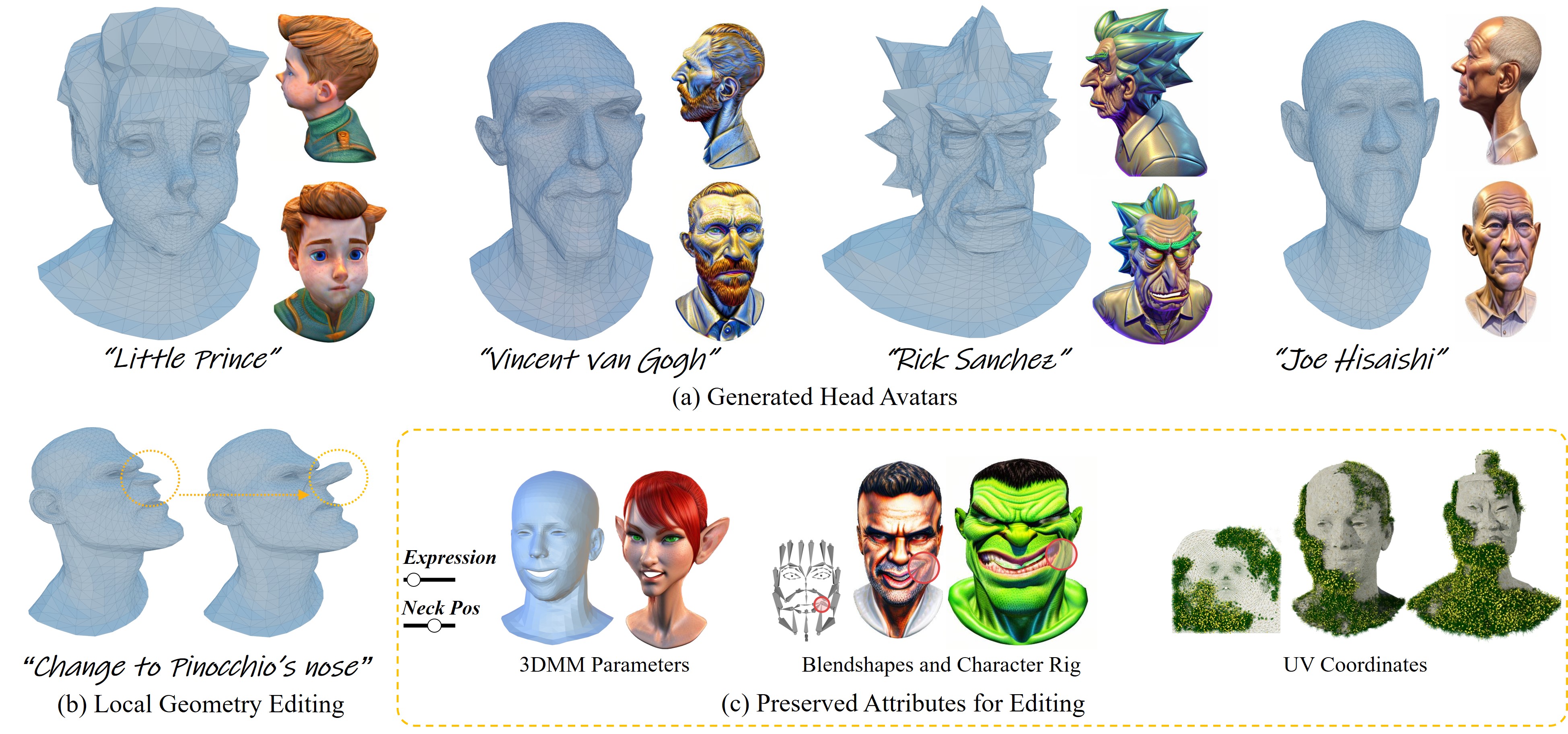}
  \captionof{figure}{Generated results and supported user editing in our framework. (a) \HeadEvolver\ can deform a template mesh to stylized head avatars under text guidance. (b) Users can use a text prompt to edit local geometry. (c) Shape attributes such as 3DMM parameters, blendshapes, and UV coordinates are preserved for downstream animation and editing.}
\label{fig:teaser}
\vspace{0.8cm}
}
\maketitle

\ificcvfinal\thispagestyle{empty}\fi

\begin{abstract}
  Current text-to-avatar methods often rely on implicit representations (e.g., NeRF, SDF, and DMTet), leading to 3D content that artists cannot easily edit and animate in graphics software. This paper introduces a novel framework for generating stylized head avatars from text guidance, which leverages locally learnable mesh deformation and 2D diffusion priors to achieve high-quality digital assets for attribute-preserving manipulation. Given a template mesh, our method represents mesh deformation with per-face Jacobians and adaptively modulates local deformation using a learnable vector field. This vector field enables anisotropic scaling while preserving the rotation of vertices, which can better express identity and geometric details. We employ landmark- and contour-based regularization terms to balance the expressiveness and plausibility of generated avatars from multiple views without relying on any specific shape prior. Our framework can generate realistic shapes and textures that can be further edited via text, while supporting seamless editing using the preserved attributes from the template mesh, such as 3DMM parameters, blendshapes, and UV coordinates. Extensive experiments demonstrate that our framework can generate diverse and expressive head avatars with high-quality meshes that artists can easily manipulate in graphics software, facilitating downstream applications such as efficient asset creation and animation with preserved attributes. Our project page is at: \href{https://www.duotun-wang.co.uk/HeadEvolver/}{https://www.duotun-wang.co.uk/HeadEvolver/}
\end{abstract}

\hypersetup{hidelinks
}

\section{Introduction}
\label{Sec:introduction}


Head avatar modeling is an important and challenging task in visual computing due to its increasing need in applications such as games~\cite{AvatarGames:Play:2007}, movies~\cite{AvatarMovie:CG:2017}, virtual reality~\cite{AvatarVR:CHI:2023}, and online education~\cite{AvatarEdu:CHI:2023}. Traditionally, creating a high-fidelity head avatar demands skilled technical artists to invest considerable time and effort into modeling and animation. 
While recent advances in AI-generated content (AIGC), particularly in text-to-3D, offer new avenues for accessible and cost-effective avatar creation, some practical gaps persist. For example, several methods~\cite{Rodin:CVPR:2023, DreamFace:TOG:2023, AvatarBooth:Arxiv:2023} used generative models~\cite{DDPM:NIPS:2020, GAN:NIPS:2020, DDIM:ICLR:2020} to support text-guided human head generation. Nevertheless, these methods suffer from acquiring high-quality and diverse datasets for training. Another set of methods~\cite{HeadSculpt:NIPS:2023, Text2control3d:Arxiv:2023, TECA:Arxiv:2023, HeadArtist:Arxiv:2023} leveraged frozen large-scale vision language models (VLMs) to create 3D head avatars. These methods use 2D priors to provide multi-view guidance without requiring 3D training data and thus produce more diverse results.

However, there is a significant gap between current text-to-avatar methods and 3D asset creation, as most methods do not effectively integrate high fidelity, user customization, and animation support.
This is mainly caused by employing implicit shape representations that cannot preserve 3D attributes essential for downstream manipulation, e.g., mesh topology, rig, and UV mapping.
For instance, HeadArtist~\cite{HeadArtist:Arxiv:2023} and HeadSculpt~\cite{HeadSculpt:NIPS:2023} leverage DMTet~\cite{DMTet:NIPS:2021} to create head avatars with 3D Morphable Models (3DMMs)~\cite{3DMMs:SGP:1999, FLAME:TOG:2017} as shape priors. While these approaches create a realistic appearance, DMTet discards statistical parameters from 3DMMs, leading to noisy meshes.

Head avatar generation using an explicit mesh representation would benefit downstream applications because of its editability and compatibility with graphics pipelines, although there is less existing research in this direction. A notable method is TADA~\cite{TADA:3DV:2024}, which incorporates the optimizations of vertex displacement and SMPL-X parameters~\cite{SMPL-X:CVPR:2019}. This strategy preserves the semantics of 3DMMs and enables animatable 3D head avatar generation. However, direct vertex displacement tends to deform meshes with self-intersected faces and noisy normals, leading to limited mesh quality and artifacts in the animation.
TextDeformer~\cite{SIGGRAPH:TextDeformer:2023} can change an input mesh smoothly and stably by deforming through Jacobians, but it cannot generate a textured appearance and lacks local geometry details.

This paper aims to create high-quality avatars that support animation and editing without reconstructing blendshapes or rigs. We propose a novel text-to-avatar framework that produces stylized head avatars through expressive and attribute-preserving mesh deformation. Given a single template mesh such as FLAME~\cite{FLAME:TOG:2017}, ICT-Face~\cite{ICTFace:Arxiv:2020}, or any manifold mesh preferably with facial semantics, our framework deforms it into a desired target shape while preserving feature correspondences such as skinny weights and UV coordinates for animation and editing, as shown in Figure~\ref{fig:teaser}(c).




The key observation is that optimizing Jacobians enables smooth global shape changes~\cite{SIGGRAPH:NJF:2022} but lacks expressive local deformations when processing gradients from text and image losses, compared to direct vertex displacements. This local control is essential for generating diverse styles, such as the elongated nose in Figure~\ref{fig:teaser}(b). Furthermore, it is often laborious to tune deformation hyper-parameters to achieve both text-guided expressiveness and shape smoothness~\cite{Largesteps:SIGGRAPH:2021}. Therefore, we propose per-triangle learnable weighting factors coupled with Jacobians to enhance fine-grained deformations, which we refer to as vector fields. The notion of vector fields was initially introduced through a Poisson-based solver~\cite{vectorField:SIGGRAPH:2004} to guide mesh deformations and reconstructions. In this study, we expand upon this by incorporating differentiable settings for 3D generation tasks. 

In order to build a complete 3D head avatar generation framework via gradient-based optimization, we employ a pre-trained diffusion model~\cite{DreamFusion:ICLR:2022} with 3D awareness by leveraging a frozen landmark-guided ControlNet~\cite{ControlNet:ICCV:2023, MediaPipe:ArXiv:2019}. Given a camera pose, we render shaded and normal images and obtain the opacity mask and the predefined vertex-indexed landmarks projected from the 3D head. As Figure~\ref{Fig:Pipeline} illustrates, the stable diffusion model accepts text prompts, rendered images, and landmarks as additional conditions from ControlNet to compute the Score Distillation Sampling (SDS) loss. In addition, to mitigate unexpected geometry distortions and self-intersecting triangles, we introduce landmark- and contour-based regularization terms. In summary, our contributions are as follows. 

\begin{itemize}
    \item We introduce a per-triangle vector field to enhance the expressiveness of mesh deformation over local regions while preserving 3D attributes.
    \item We propose a diffusion-based framework with landmark- and contour-based regularization terms for high-quality text-to-avatar mesh generation.

    \item Extensive experiments demonstrate that our method can generate stylized 3D head avatars supporting asset creation with interactive editing and animation.
\end{itemize}

\section{Related Work}
\label{Sec:relatedwork}
Recently, there has been rapid research progress in digital human modeling and text-to-3D content generation. Our work centers on text-guided synthesis of head avatars employing an explicit mesh representation, aiming to fulfill three criteria outlined in Table~\ref{tab: features}. In addition, our framework generates mesh models that are compatible with graphic pipelines for direct manipulation in 3D software.


\begin{table}
\scriptsize
\caption{Compared to other text-to-avatar approaches, our method preserves predefined 3D attributes (e.g., vertex-indexed landmarks, rig, blendshapes). It accommodates any template mesh independent of parametric models and operates solely on gradient-based optimizations without training data or fine-tuning.}
\vspace{2mm}
\centering
\begin{tabular}{c c c c}
     \hline 
      & Attribute& Non-reliance on & No Training Data\\
      & Preservation & Parametric Models & nor Fine-tuning \\
     \hline
     HeadScuplt~\cite{HeadSculpt:NIPS:2023} & \xmark & \cmark  & \xmark\\
     Fantasia3D~\cite{Fantasia3D:ICCV:2023} & \xmark & \cmark  & \cmark\\
     TADA~\cite{TADA:3DV:2024} & \cmark & \xmark  & \cmark\\
     DreamFace~\cite{DreamFace:TOG:2023} & \cmark & \xmark  & \xmark\\
     Ours & \cmark & \cmark  & \cmark\\
     \hline
\end{tabular}
\label{tab: features}
\end{table}

\subsection{Text-to-3D Avatar Generation}
\label{related work:text-to-3D}
The success of text-driven generative models has sparked a wide range of research in 3D content synthesis. These methods are based on different geometry representations such as NeRF~\cite{Magic3D:CVPR:2023, ProlificDreamer:NIPS:2023}, tri-plane~\cite{TriplaneDiffusion:CVPR:2023}, mesh\cite{Fantasia3D:ICCV:2023}, and Gaussian Splatting~\cite{DreamGaussian:Arxiv:2023, Luciddreamer:Arxiv:2023}. In text-to-avatar generation, many efforts have been made to develop 3D generative models~\cite{Rodin:CVPR:2023, DreamFace:TOG:2023, TG-3DFace:ICCV:2023, styleavatar3d:Arxiv:2023} typically based on GANs~\cite{GAN:NIPS:2014} or diffusion models~\cite{DDPM:NIPS:2020}. These methods often demand extensive 3D head data for training, incurring high costs. Moreover, they face challenges due to limited 3D datasets for creating diverse head avatars matching text prompts.

To bypass the extensive training process of 3D generative models and insufficient 3D data, recent research~\cite{Avatarclip:SIGGRAPH:2022, Avatarstudio:Arxiv:2023} has made remarkable advances by harnessing the capabilities of vision-language models (e.g., CLIP~\cite{CLIP:CORR:2021} and SDS~\cite{DreamFusion:ICLR:2022}) to generate static human avatars from text prompts. DreamWaltz~\cite{Dreamwaltz:NIPS:2023} aims to enhance the realism of avatar animations with pose-guided ControlNet~\cite{ControlNet:ICCV:2023}. Similarly, DreamHuman~\cite{Dreamhuman:NIPS:2024} employs a signed distance field conditioned on pose and shape parameters to empower animations with NeRF. However, extracting, editing, and animating an explicit mesh from these works are not straightforward~\cite{Dreamwaltz:NIPS:2023} (e.g., use marching cube or tetrahedra algorithms) since NeRF and DMTet~\cite{DMTet:NIPS:2021}, for instance, represent shapes through network weights. Meanwhile, the issue of poor mesh quality is often concealed by texture, as the geometry and texture optimizations are fully mixed. Recent works~\cite{Fantasia3D:ICCV:2023, HeadSculpt:NIPS:2023, HeadArtist:Arxiv:2023} disentangle the learning of geometry and texture to improve the avatar mesh quality. Our work also leverages differentiable rendering, stable diffusion, and a hybrid optimization process for mesh and texture synthesis, specifically emphasizing the deformation of an explicit geometry template through semantic guidance instead of generation from scratch or using implicit representations.

\subsection{Learning-Based Mesh Optimization}  
Parametric mesh templates have been a prominent research focus for human avatar reconstruction and generation tasks, as these models are well-suited in a deep learning context. AvatarClip~\cite{AvatarCLIP:TOG:2022} and CLIP-Face~\cite{ClipFace:SIGGRAPH:2023} employ parametric models such as SMPL-X~\cite{SMPL-X:CVPR:2019} and FLAME~\cite{FLAME:TOG:2017} as geometry prior to produce text-aligned avatars. DreamAvatar~\cite{Dreamavatar:Arxiv:2023} extracts the shape parameters of SMPL as a 3D prior to learn a NeRF-based color field. HeadSculpt~\cite{HeadSculpt:NIPS:2023} and HeadArtist~\cite{HeadArtist:Arxiv:2023} utilize the FLAME template as canonical shape input and extract its facial landmark to guide avatar generations through DMTet. However, meshes extracted from implicit representations tend to lose the semantics of 3DMMs~\cite{3DMMs:SGP:1999}, posing challenges for downstream applications such as interactive editing and animation.

From the perspective of geometry editing through explicit deformation, recent works propose data-driven approaches~\cite{FastDeform:TOG:2020, SIGGRAPH:NJF:2022} to predict realistic deformation and some methods focus on learning deformation over one template mesh through tuning per-vertex coordinates~\cite{3DStyleNet:ICCV:2021, DeepEmulator:CVPR:2021, UnpairedShape:Graph:2018}. TADA~\cite{TADA:3DV:2024} directly optimizes vertex displacements along with the SMPL-X's pose and shape parameters and achieves animatable avatars. However, this characteristic is highly dependent on the quality of 3DMMs. Straightly optimization of non-convex mesh structures often converges to undesirable local minima, resulting in noticeable artifacts on surface details. TextDeformer~\cite{SIGGRAPH:TextDeformer:2023} initially integrates CLIP into a learning-based deformation pipeline to address the scarcity of 3D datasets that pair shapes with captions. Our approach extends this by employing vector fields to optimize local semantic deformations on any template mesh, independent of its shape, expression, or blendshapes~\cite{3DMMs:SGP:1999}. Our goal is to enhance identity expressiveness guided by text input. The resulting head avatars can be animated using the original rigging from the source mesh and seamlessly integrated into existing 3D design and animation workflows.

\section{Methodology}
\label{Sec:method}

\begin{figure*}[tbh]
\includegraphics[width=0.98\textwidth]{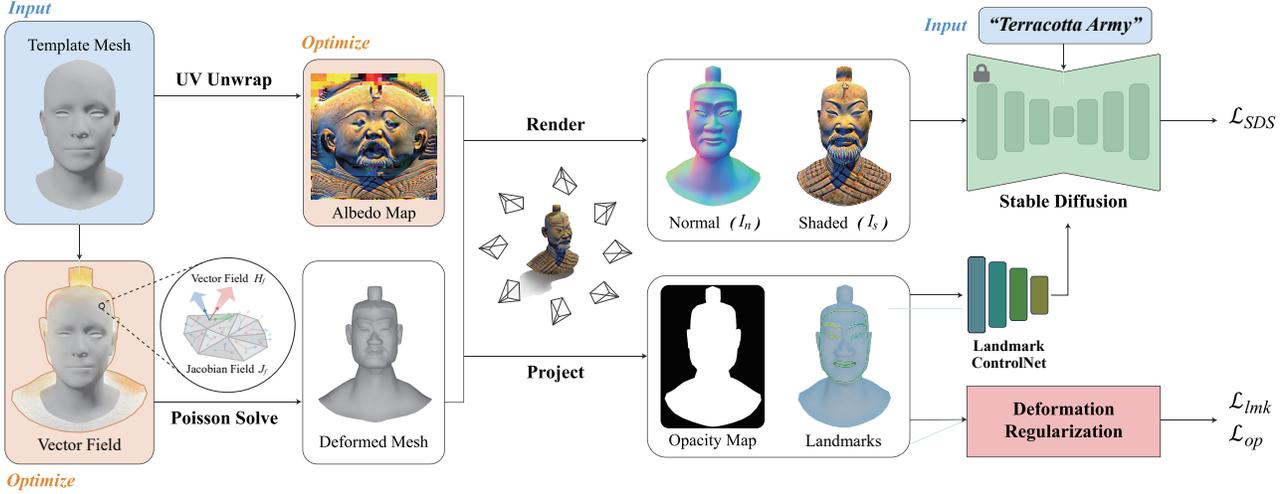}
\caption{Framework overview. We deform a template mesh by optimizing per-triangle vector fields guided by a text prompt. Rendered normal and RGB images coupled with MediaPipe landmarks are fed into a diffusion model to compute respective losses. Our regularization of Jacobians controls the fidelity and semantics of facial features that conform to text guidance.}
\label{Fig:Pipeline}
\end{figure*}

In this section, we describe the details of our generation pipeline, as shown in Figure~\ref{Fig:Pipeline}. Given a source template mesh, we jointly optimize the mesh deformation and an albedo map guided by a text prompt. Our approach is not limited to a specific parametric model (e.g., FLAME), and Figures~\ref{Fig:Fullbody Demonstration} and~\ref{fig:obj_eval} show the generalization of our approach. 


\begin{figure}[tbh]
\includegraphics[width=0.48\textwidth]{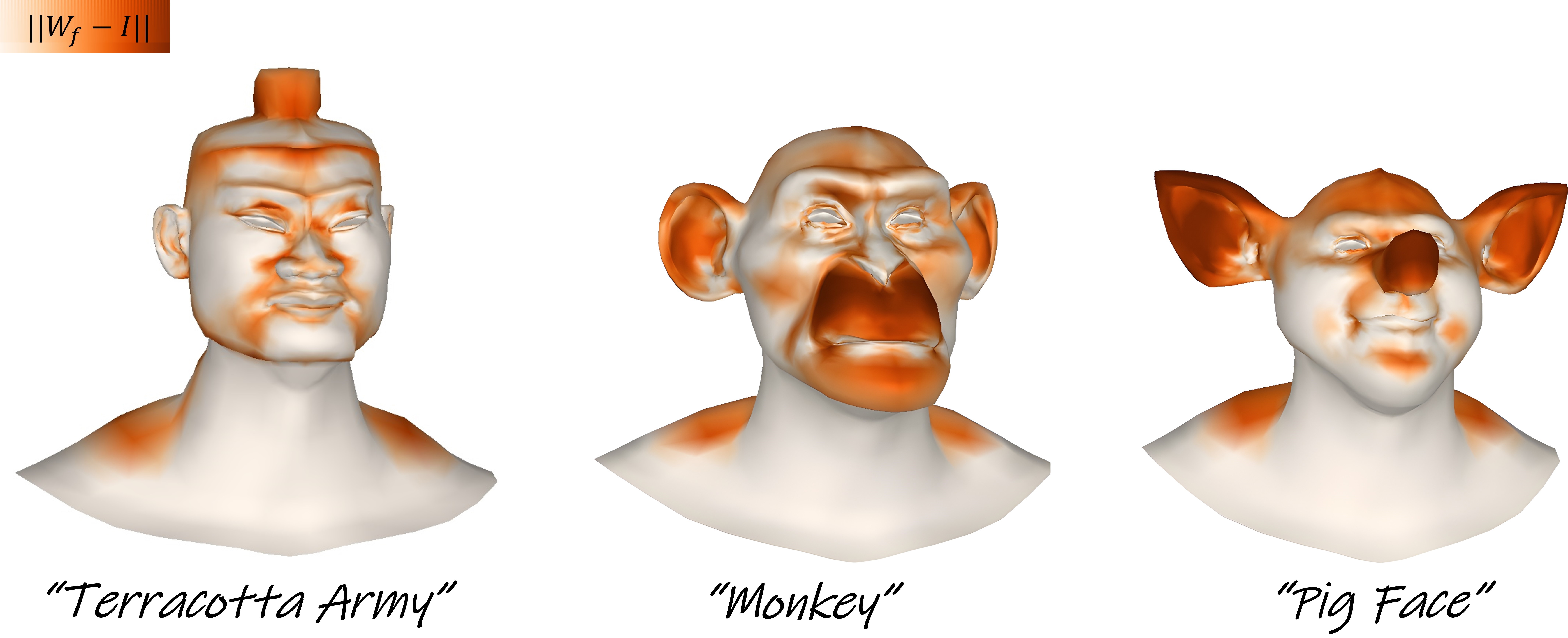}
\caption{Visualization of vector field $H_f$ through $W_f$. Orange colors indicate strong deformations and $W_f=I$ refers to no deformation enhancement.}
\label{Fig:weight visualization}
\end{figure}


\subsection{Mesh Deformation through Vector Fields}
\label{Sec:deformation method}
Let $\mathcal{M}=(\mathcal{V}, \mathcal{F})$ denote a source template mesh which consists of a set of vertices $\mathcal{V} \in \mathbb{R}^{n\times 3}$ and faces $\mathcal{F} \in \mathbb{Z}^{m\times 3}$. For the optimization strategy of mesh deformation, directly applying displacement to each vertex can result in severe self-interactions of triangle faces since it is susceptible to localized noisy gradients from pixel-level losses. Instead, we parameterize deformation mapping $\Phi: \mathbb{R}^{n\times 3} \to \mathbb{R}^{n\times 3}$ by employing Jacobian fields~\cite{SIGGRAPH:NJF:2022} $J=\{J_f | f \in \mathcal{F}\}$ that represents the scaling and rotation of each mesh face. A dual representation of $\mathcal{M}$ can be defined as a stack of per-face Jacobians $J_f \in \mathbb{R}^{3\times 3}$ where  
\begin{equation}
J_f = \nabla_f\mathcal{V}, 
\end{equation}
and $\nabla_f$ is the gradient operator of triangle $f$. Conversely, we optimize each vertex's position from per-face Jacobian.

After obtaining optimized Jacobians in each iteration, we can compute the deformation mapping $\Phi$ over a set of vertices conforming to the source mesh topology by solving a Poisson problem, i.e.,
\begin{equation}
\Phi^{\ast} = \text{arg} \min_{\Phi}\sum_{f \in \mathcal{F}}|f| \lVert \nabla_{f}(\Phi)- J_{f}\rVert^{2}_2,
\label{eq: deformation mapping}
\end{equation}
where $\nabla_{f}(\Phi)$ is the Jacobian of $\Phi$ at a triangle face $f$ and $|f|$ is the area of the face. Thus, the deformation map $\Phi$ over the input template mesh is indirectly optimized through Jacobians $J_f$ in the least square sense. The solution can be obtained by using a differentiable solver~\cite{SIGGRAPH:NJF:2022, SIGGRAPH:TextDeformer:2023}: 
\begin{equation}
\Phi^{\ast} = L^{-1}\nabla^{T}\mathcal{A}J,
\label{eq: phi_star}
\end{equation}
where $\nabla$ is a stack of per-face gradient operators, $\mathcal{A} \in \mathbb{R}^{3m \times 3m}$ is the mass matrix of $\mathcal{M}$, and $L \in \mathbb{R}^{3n \times 3n}$ is the cotangent Laplacian of $\mathcal{M}$. This learnable shape representation based on Jacobians is robust to noisy gradients. It can achieve better geometry quality (e.g., fewer self-intersections and noisy normals) than direct vertex displacements, which is further justified in Section~\ref{Sec:Experiments}.

However, during our experiment, we observed that the Poisson solver applied in Equation~\ref{eq: deformation mapping} can impose strong geometry constraints for Jacobians to reach the global-coherence mesh deformation~\cite{SIGGRAPH:TextDeformer:2023}, which may lose fine-grain details for local shape features and compromise the identity of the generated head mesh (Figure~\ref{Fig:Weighted Jacobian Ablation}). A straightforward approach is to increase the learning rate for Jacobians to enhance their sensitivities to propagated gradients. This amplifies the rotation and scaling of each face, resulting in stronger deformations. However, intensified rotations raise the likelihood of inverted triangles~\cite{InjectiveMapping:SGP:2013}, leading to an unstable optimization process and a poor-quality mesh. To address the above concerns, we introduce a learnable per-face vector field $H = \{H_f|f \in \mathcal{F}\}$ to enhance the deformation expressiveness of Jacobian field $J$, i.e.,
\begin{equation}
\begin{split}
W_f &= diag(w_x, w_y, w_z), \\
H_f &= W_f J_f. 
\end{split}
\end{equation}
By substituting $H$ into Equation~\ref{eq: phi_star}, the deformation mapping could be computed by 
\begin{equation}
\Phi^{\ast} = L^{-1}\nabla^{T}\mathcal{A}H.
\end{equation}
The effect of $W=\{W_f|f\in\mathcal{F}\}$ is to control the anisotropic scaling (e.g., stretching and shrinking in different directions) on each face.  As shown in Figure~\ref{Fig:weight visualization}, it can adaptively learn the anisotropic scaling on corresponding triangular faces that can represent the character's features. Consequently, the per-face vector field $H$ prioritizes the Poisson solver for more effective head avatar identity expression than using only Jacobians (Section~\ref{Ablation}).

\subsection{Text-Guided 3D Synthesis}  
We aim to build a text-guided head avatar generation framework based on differentiable mesh deformation and texture generation. Moti.vated by recent success in large-scale VLMs and text-to-3D literature discussed in Section~\ref{related work:text-to-3D}, we utilize a powerful 2D diffusion prior (i.e., SDS~\cite{DreamFusion:ICLR:2022}) to direct deformations by scoring the rendering realism of the generated shape in our framework. 

Specifically, by following the procedure described in Section~\ref{Sec:deformation method}, we denote $\mathcal{V^*}$ as the set of deformed vertices from $J$. Then, by using a differentiable render $\mathbf{R}$, we render $\mathcal{M^*}=(\mathcal{V}^*, \mathcal{F})$ from a viewpoint delineated by camera extrinsic parameter $C$, which produces an image $I=\mathbf{R}(\mathcal{M^*}, C)$. The SDS loss is leveraged to alter the geometry and color respectively, that is:
\begin{equation}
\footnotesize
\nabla_{J} \mathcal{L}_{\mathrm{SDS}}(\epsilon, I)=\mathbb{E}_{t,\epsilon}\left[w(t)(\epsilon_{\theta}(z_{t};y,C, t)-\epsilon)\frac{\partial{I}}{\partial{J}}\right],
\end{equation}
where $\epsilon \sim \mathcal{N}(0, I)$, $\epsilon_{\theta}$ is the pre-trained diffusion priors with the network parameter $\theta$, $w(t)$ is a weight coefficient related to timestep $t \sim \mathcal{U}(0,1)$, $y$ is the text input condition, $z_{t}$ denotes a latent embedding of $I$ perturbed with noises at time step $t$, and $C$ is the MediaPipe~\cite{MediaPipe:ArXiv:2019} facial landmark map. Rendered image $I$ are obtained and encoded from the normal $I_{n}$ and shaded $I_{s}$, as seen in Figure~\ref{Fig:Pipeline}. 


To ensure deformation mapping from propagated gradients and the Poisson solver can conform to the facial topology and reasonable semantics of the source mesh, we further develop two regularization terms, aiming to preserve 3D attributes of generated head avatars (Figure~\ref{Fig:semantics-coherence Demonstration}).

\textbf{Landmark regularization.} The landmark regularization measures the difference between predefined $N$ 3D face landmarks $\mathbf{k}_i \in \mathbb{R}^{3}$ on the canonical template mesh $\mathcal{M}$ and the corresponding ones on the deformed mesh $\mathbf{k}'_i \in \mathbb{R}^{3}$ after projection. The landmark regularization loss is defined as $\mathcal{L}_{lmk} = \frac{1}{N} \sum_{i=1}^{N} ||\mathbf{k}_i - \mathbf{k}'_i||^{2}_{2}$.

\textbf{Evolving contour-based regularization.}
Inspired by AvatarCraft~\cite{AvatarCraft:ICCV:2023}, we incorporate an opacity map $O$ to preserve a plausible head shape while minimizing excessive or undesired vertex displacement. To balance diverse geometry shapes and constraints from severe deformations, the source template mesh is periodically updated from deformed mesh every 300 steps~\cite{SEEAvatar:Arxiv:2023}. The pixel-level loss is parameterized as $\mathcal{L}_{op}=\frac{1}{hw}\sum_{h,w}||O - O'||^{2}_{2}$, where $h$ and $w$ represent the height and width of the opacity mask of the template mesh $O$ and the optimized head avatar $O'$. The total loss is defined as follows:
\begin{equation} 
\mathcal{L}= \lambda_{1}\nabla_{\Phi}\mathcal{L}_{\mathrm{SDS}}+
\lambda_{2} \mathcal{L}_{lmk} + \lambda_{3} \mathcal{L}_{op},
\label{eq: overall loss}
\end{equation}
where $\lambda_{1}=1$, $\lambda_{2}=200$, and $\lambda_{3}=250$ are the hyper-parameters in our experiment settings.

\section{Experiments}
\label{Sec:Experiments}

In this section, we conduct experiments to evaluate the efficacy of our method both quantitatively and qualitatively for text-to-avatar creation. We then present two ablation studies that validate the significance of our key designs for vector fields and regularization terms in Section~\ref{Ablation}. 

\subsection{Implementation Details}
\label{implementation}
For geometry and texture generation, we render normal and shaded images at a resolution of $1024\times 1024$ pixels and feed them into the Realistic Vision 5.1 (RV 5.1)~\cite{RV51}. Compared with Stable Diffusion 2.1~\cite{StableDiffusion:Arxiv:2021}, we find that RV 5.1 supports a more appealing avatar appearance. The multi-face Janus and texture misalignment problems are further alleviated by introducing ControlNetMediaPipeFace~\cite{ControlNet:ICCV:2023}. The generation process runs on a single NVIDIA RTX $4090$ GPU with $20,000$ iterations per avatar, which takes around $30$ minutes. Negative prompts are leveraged to enhance the realism and details of the texture, e.g., ``unrealistic,'' ``blurry,'' ``low quality,'' ``saturated,'' ``low contrast,'' etc. Along with the proposed regularization terms, we enforce symmetry along the y-axis in each iteration to ensure a reasonable face appearance. Following established practices in TADA and DreamFace, the eyeball mesh is removed and can be manually reintegrated after the generation.

\begin{figure}[!htb]
    \centering
\includegraphics[width=0.475\textwidth]{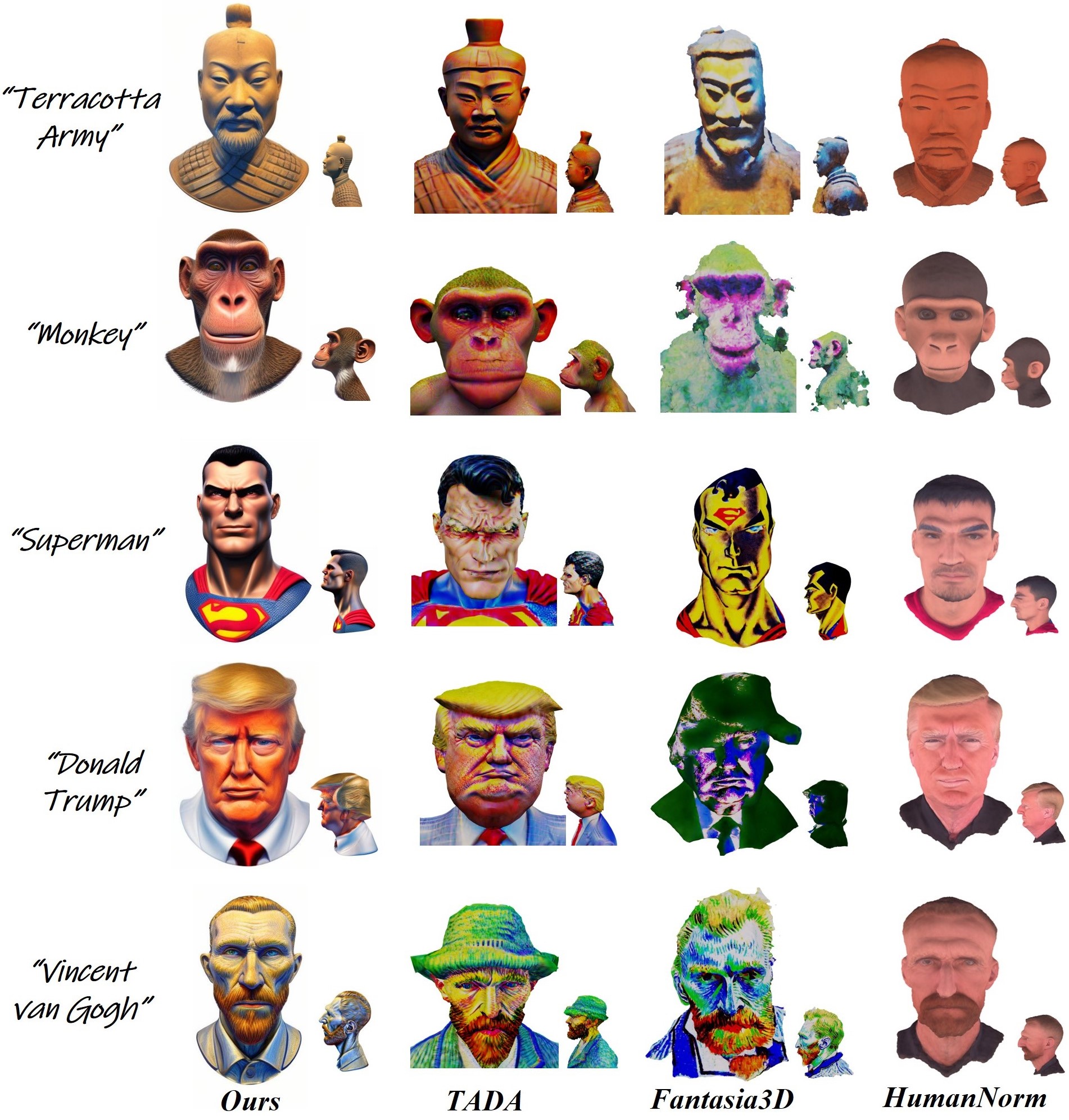}
\caption{Qualitative appearance comparisons. Our method can produce diverse textured avatars.}
\label{Fig:Qualitative Comparison Appearance}
\end{figure}


\begin{figure*}[!htb]
\includegraphics[width=1.0\textwidth]{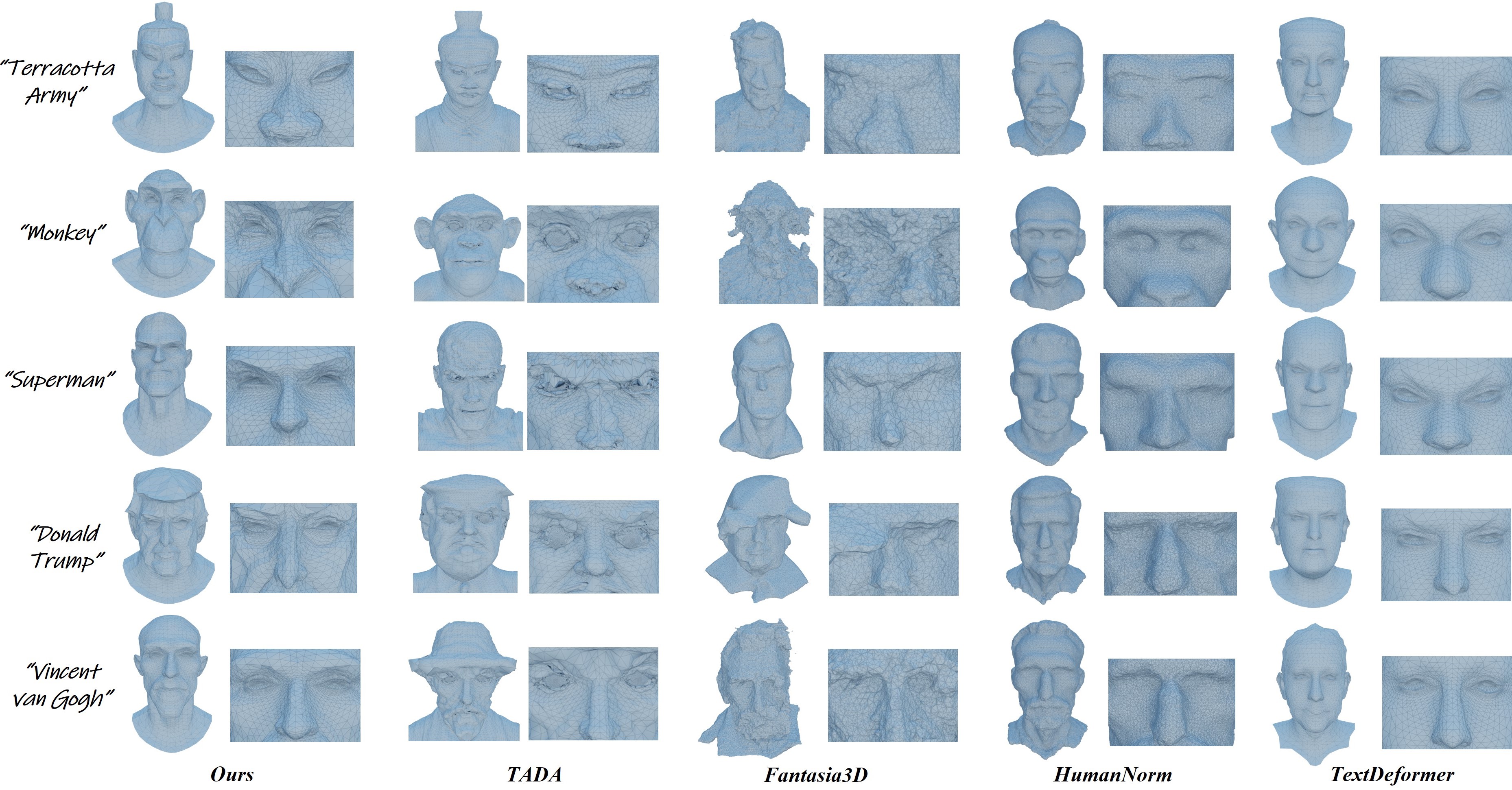}
\caption{Qualitative geometry comparisons. Our method effectively preserves the topology and semantics of the input template mesh, resulting in high-quality mesh models for smooth manipulations.}
\label{Fig:Qualitative Comparison Geometry}
\end{figure*}


\subsection{Qualitative Evaluation}
Our evaluation compares two explicit mesh manipulation methods (i.e., TADA~\cite{TADA:3DV:2024} and TextDeformer~\cite{SIGGRAPH:TextDeformer:2023}), and two implicit methods (i.e., Fantasia3D~\cite{Fantasia3D:ICCV:2023} and HumanNorm~\cite{HumanNorm:CVPR:2024}) as baselines. While our framework focuses on high-quality mesh topologies, we also compare rendering appearance with recent non-mesh-dependent methods~\cite{HeadArtist:Arxiv:2023, HeadSculpt:NIPS:2023, headstudio:arxiv:2024} in supplemental material (Figure~\ref{fig:comp_implicit}). We utilized FLAME~\cite{FLAME:TOG:2017} as a unified input template mesh to conduct the experiments. We present the visual comparison results in Figures~\ref{Fig:Qualitative Comparison Appearance} and~\ref{Fig:Qualitative Comparison Geometry}. TADA and Fantasia3D show detail-preserving geometry with 3D head priors but still undergo geometric artifacts in dense areas like eye regions, which are less noticeable in textured rendering. HumanNorm underperforms in humanoid cases, e.g., ``Terracotta Army'' and ``Monkey.'' Compared to TextDeformer, our approach exhibits similar mesh quality and amplifies facial characteristics following text specifications.

In summary, our framework generates expressive head meshes that inherit facial attributes. The texture generation maintains decent geometry consistency, facilitating usage in graphics applications (Figures~\ref{fig:teaser}(c) and~\ref{Fig:Qualitative Comparison Appearance}).

\subsection{Quantitative Evaluation}
We quantitatively evaluated our framework for generation consistency with text input and generation quality.

\textbf{Generation Consistency with Text.} We first evaluated the relevance of generated results to the corresponding text descriptions by calculating the CLIP score~\cite{CLIP:CORR:2021}. We rendered 20 distinct views and computed the average scores respectively, as shown in Table~\ref{tab:quantitative comp}. Our approach achieves the highest scores compared to the baseline methods.

\begin{table}[htb]
\caption{Quantitative comparison of state-of-the-art methods. The geometry CLIP score is measured on the shaded images rendered with
uniform albedo colors~\cite{Richdreamer:CVPR:2024}, the appearance CLIP score is evaluated on the images rendered with textures, and the self-intersection is quantified as the ratio of self-intersected mesh faces to the total number of mesh faces.}
\footnotesize   
\vspace{2mm}
\begin{tabular}{c c c c}
         \hline
           & Geometry  & Appearance  & Mesh \\
         & CLIP Score $\uparrow$ &  CLIP Score $\uparrow$ & Self-Intersection $\downarrow$\\
         \hline 
         Fantasia3D & $0.1815$ & $0.2718$  & - \\
         HumanNorm & $0.2443$ & $0.3031$  & - \\
         TextDeformer & $0.2462$  & -  & $2.19\%$ \\
         TADA &  $0.2475$ & $0.2931$  & $14.22\%$ \\
         Ours & \textbf{0.2538} & \textbf{0.3089} &  \textbf{2.08\%}\\
         \hline
\end{tabular}
\label{tab:quantitative comp}
\end{table}

\begin{table}[htb]
\caption{User evaluation of generated head avatars.}
\centering
\footnotesize
\vspace{1mm}
    \begin{tabular}[width=1.0\textwidth]{c c c c}
         \hline 
         User & Geometry &  Texture  & Consistency \\
         Preference $\uparrow$  & Quality & Quality & with Text \\
         \hline 
         Fantasia3D & $0.4\%$ & $2.0\%$ & $1.9\%$ \\
         HumanNorm & $5.6\%$ & $22.7\%$ & $6.7\%$ \\
         TextDeformer & $24.2\%$ & - & $2.3\%$ \\
         TADA & $3.9\%$ & $13.1\%$ & $22.2\%$ \\
         Ours & \textbf{65.9\%} &\textbf{62.2\%} & \textbf{66.9\%} \\
         \hline
    \end{tabular}
\label{tab:user comp}
\end{table}

\textbf{Mesh Quality.} We evaluate mesh quality through self-intersection. Since the mesh from Fantasia3D and HumanNorm are extracted through the marching cubes algorithm~\cite{MC:SIGGRAPH:1987}, the produced triangles have no self-intersections. Therefore, we only compared our method with TADA and TextDeformer. Unlike our method, direct vertex displacement tends to converge to a local minimum and disregards the triangulation of the mesh.

\begin{figure}[!tbh]
\includegraphics[width=0.495\textwidth]{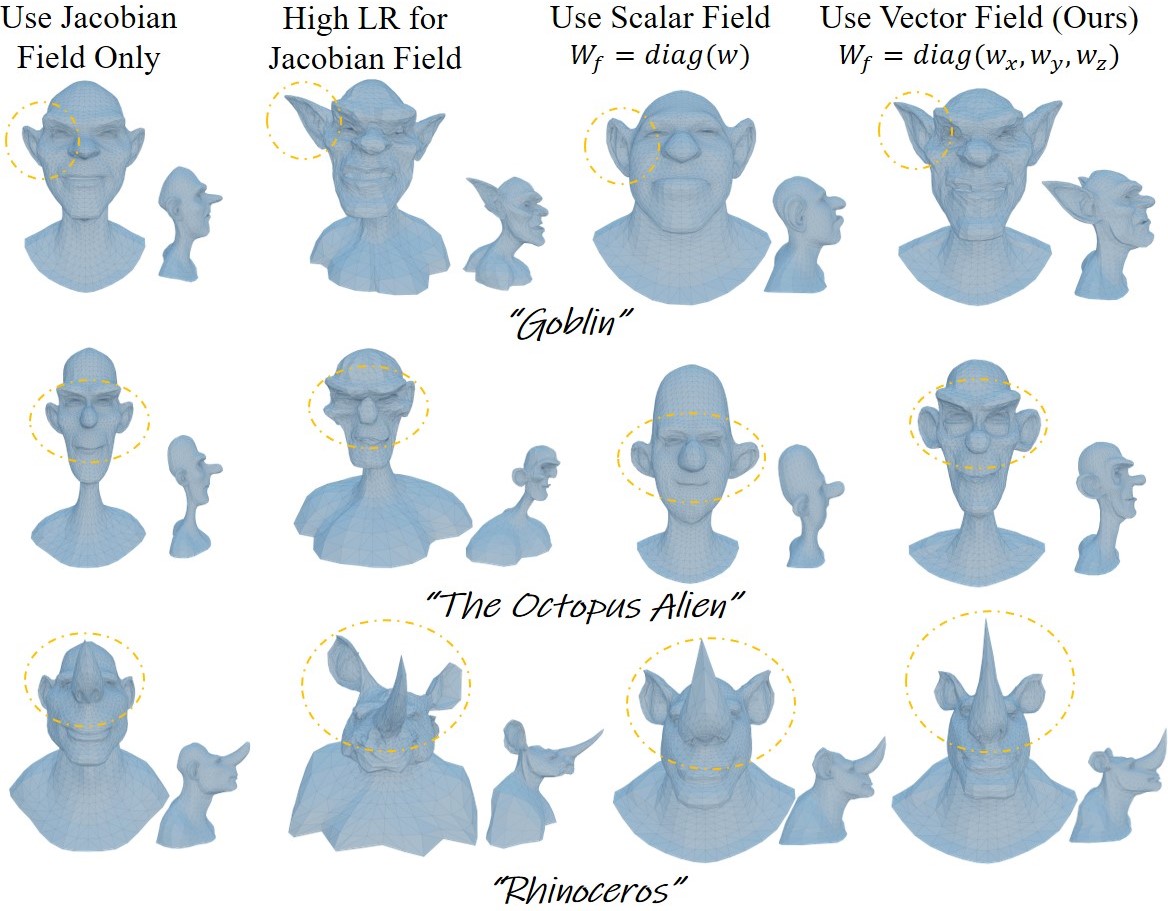}
\caption{Ablation study on the vector fields. The per-triangle $H$ can enhance the identity and character features of the avatar.}
\label{Fig:Weighted Jacobian Ablation}
\end{figure}

\begin{figure}[!tbh]
\includegraphics[width=0.49\textwidth]{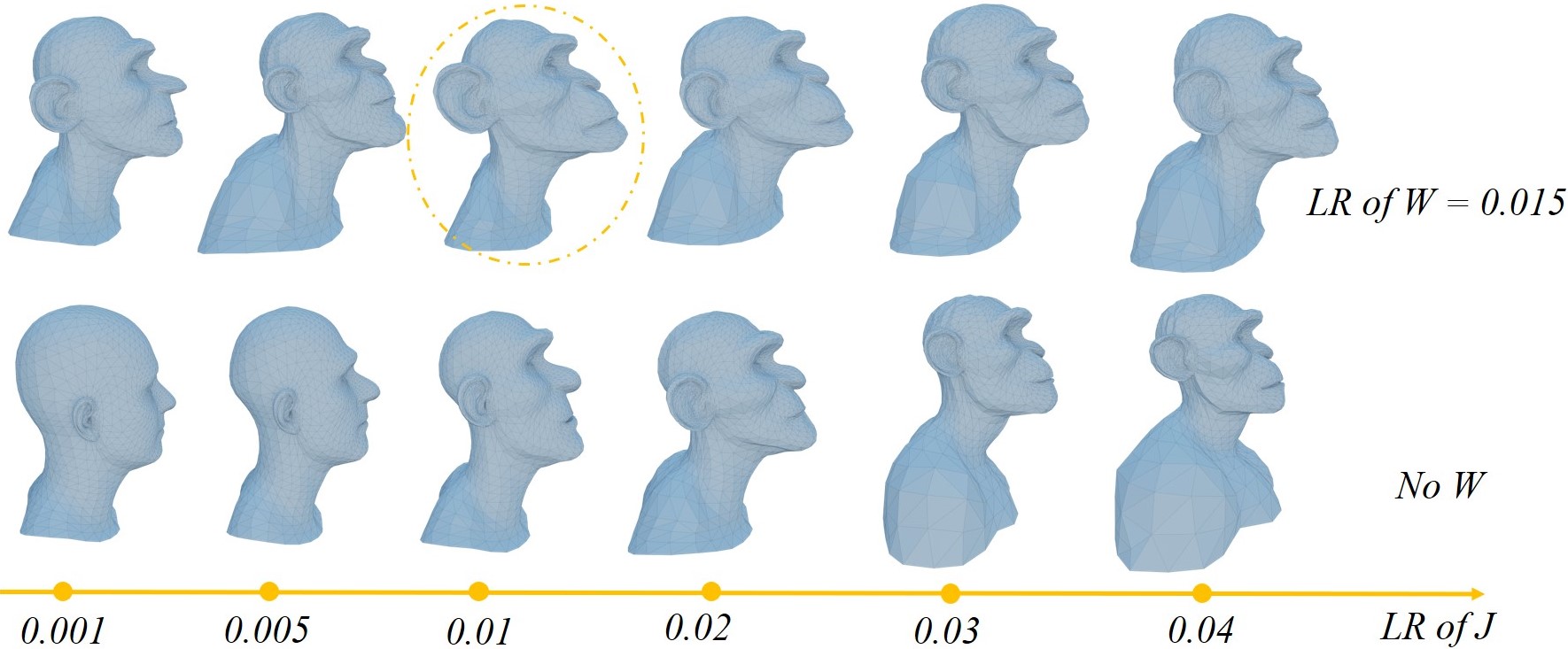}
\caption{Learning rate (LR) tuning. Supported by our experiments, adding per-triangle $H$ through $W$ can enhance the identity and character features of the avatar from the text prompt.}
\label{Fig: Jacobian Learning Rate}
\centering
\includegraphics[width=0.47\textwidth]{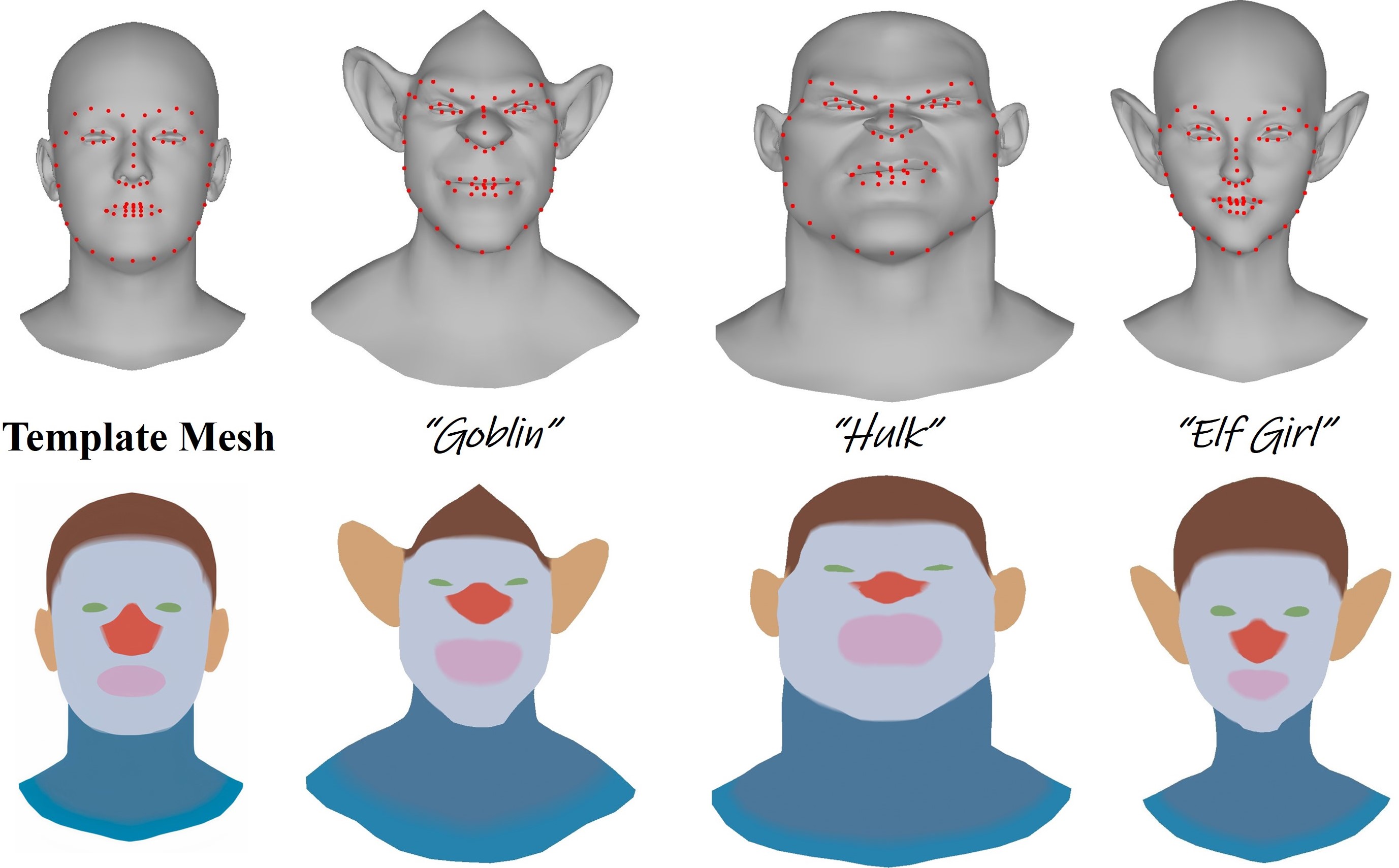}
\caption{Preserved semantic consistency (e.g., 3D landmarks and dense segmentation) with our deformation-based framework.} 
\label{Fig:semantics-coherence Demonstration}
\end{figure}

\textbf{User Study.} We conducted a user study to evaluate the robustness and expressiveness of our method with 18 distinct text prompts, 6 of which are from TADA. We used a Google Form to assess 1) geometry quality, 2) texture quality, and 3) consistency with text. We recruited 46 participants, of whom 26 are graduate students majoring in computer graphics and vision, and 20 are company employees specializing in AI content generation. In this form, the participants were instructed to choose the preferred renderings of head avatars from different methods in randomized order, as shown in Table~\ref{tab:user comp}. The results show that participants preferred our method by a significant margin (over $60\%$).




\begin{figure}[!tbh]
 \centering
\includegraphics[width=0.45\textwidth]{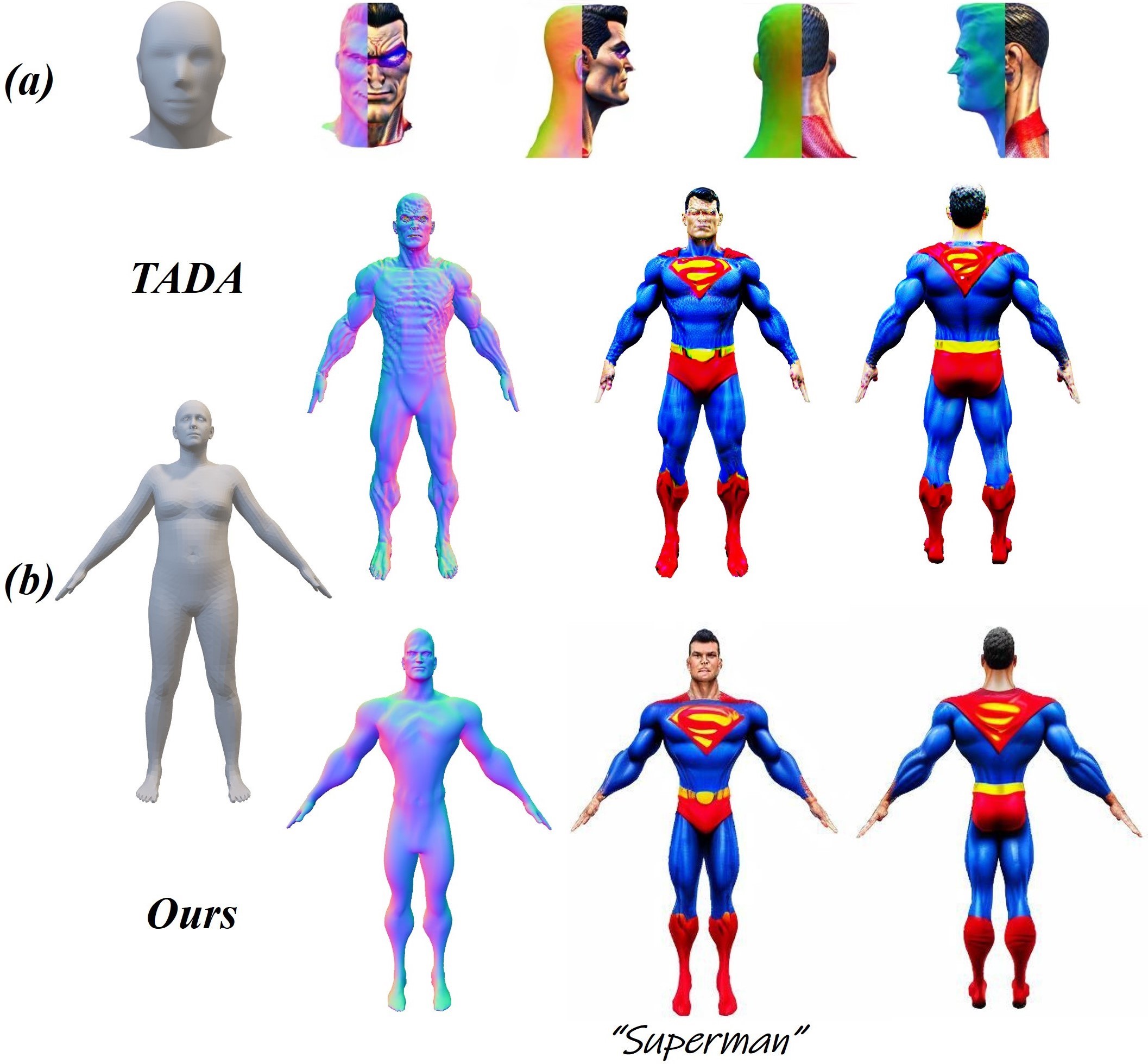}
\caption{Avatar generation with various template mesh inputs. The meshes from (a) an arbitrary head model and (b) SMPL-X~\cite{SMPL-X:CVPR:2019} are deformed to align text descriptions respectively.}
\label{Fig:Fullbody Demonstration}
\includegraphics[width=0.45\textwidth]{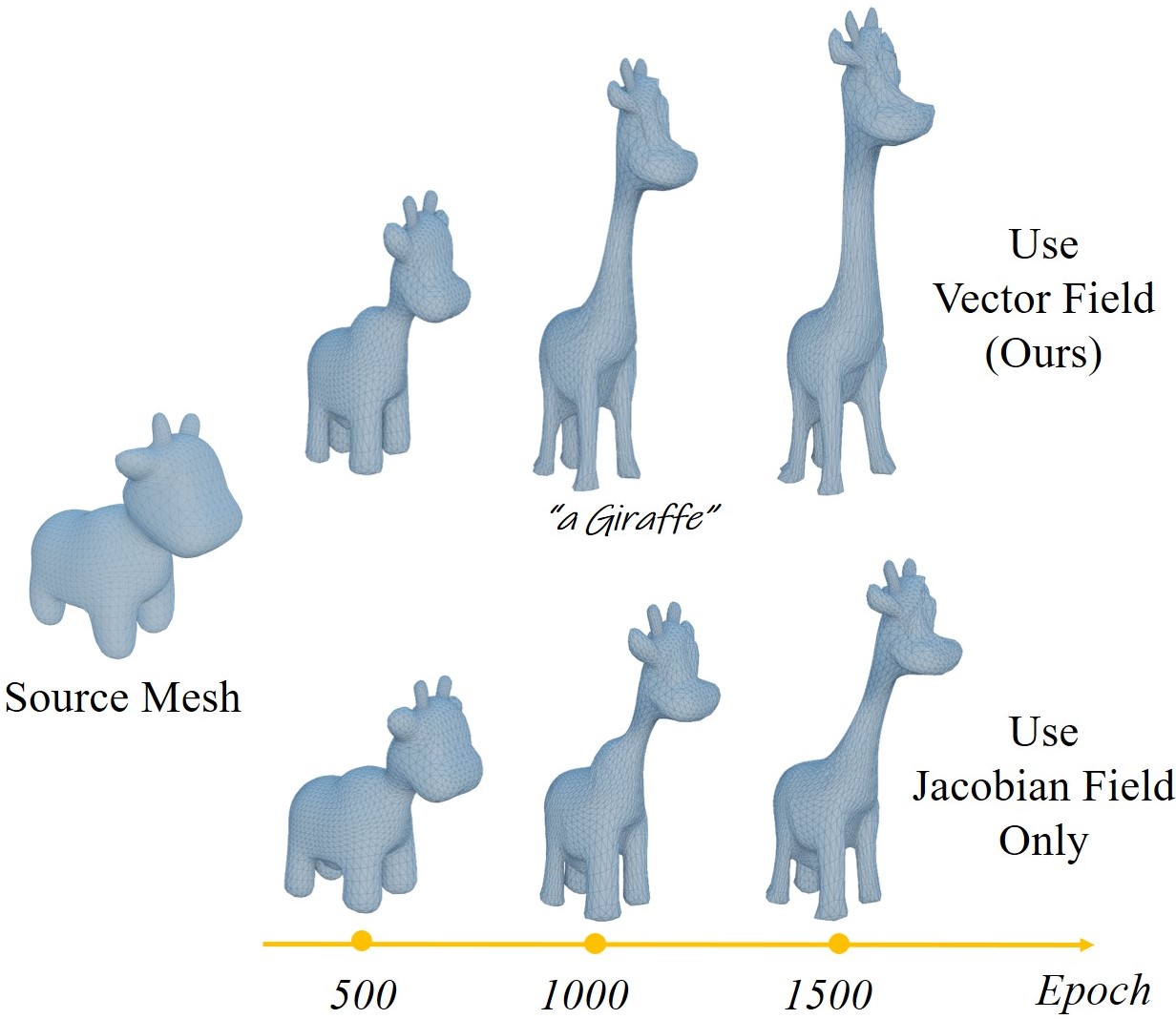}
\caption{Generalization experiments. We implement vector fields in TextDeformer~\cite{SIGGRAPH:TextDeformer:2023} and find that our method accelerates mesh optimizations to approach text guidance in early epochs.}
\label{fig:obj_eval} 

\end{figure}

\subsection{Ablation Study}
\label{Ablation}
\textbf{Effect of Vector field.} We conducted three sets of qualitative comparisons to demonstrate the effectiveness of the per-triangle vector field $H$. As shown in Figure~\ref{Fig:Weighted Jacobian Ablation}, using Jacobian $J$ alone tends to overly smooth the gradients, leading to a loss of distinctive character features. Increasing $J$'s learning rate emphasizes character identities but distorts meshes and causes self-intersections. Additionally, our experiments introduce the scalar field $S$, which represents the isotropic scaling by setting diagonal elements of each $H_i$ to equal values. The results indicate that anisotropic scaling guided through $H$ generates avatars bearing more geometry details and character features than those through $S$ (e.g., the eyebrow of "Octopus Alien" and the ear of "Goblin"). Furthermore, we demonstrate the tuning process of the LR to evaluate that the addition of $H$ effectively enhances the stability and expressiveness of deformations when the LR of $J$ is set in the reasonable range (e.g., \textbf{between 0.01 to 0.025}). In Figure~\ref{Fig: Jacobian Learning Rate}, we provide the best LR settings, which are $0.015$ for $H$ and 0.01 for $J$. We also demonstrate the versatility of the vector field by applying it to the object generation framework (Figure~\ref{fig:obj_eval}), observing that the use of $H$ accelerates optimization convergence to desired deformation states compared to using Jacobians alone.

\begin{figure}
\includegraphics[width=0.46\textwidth]{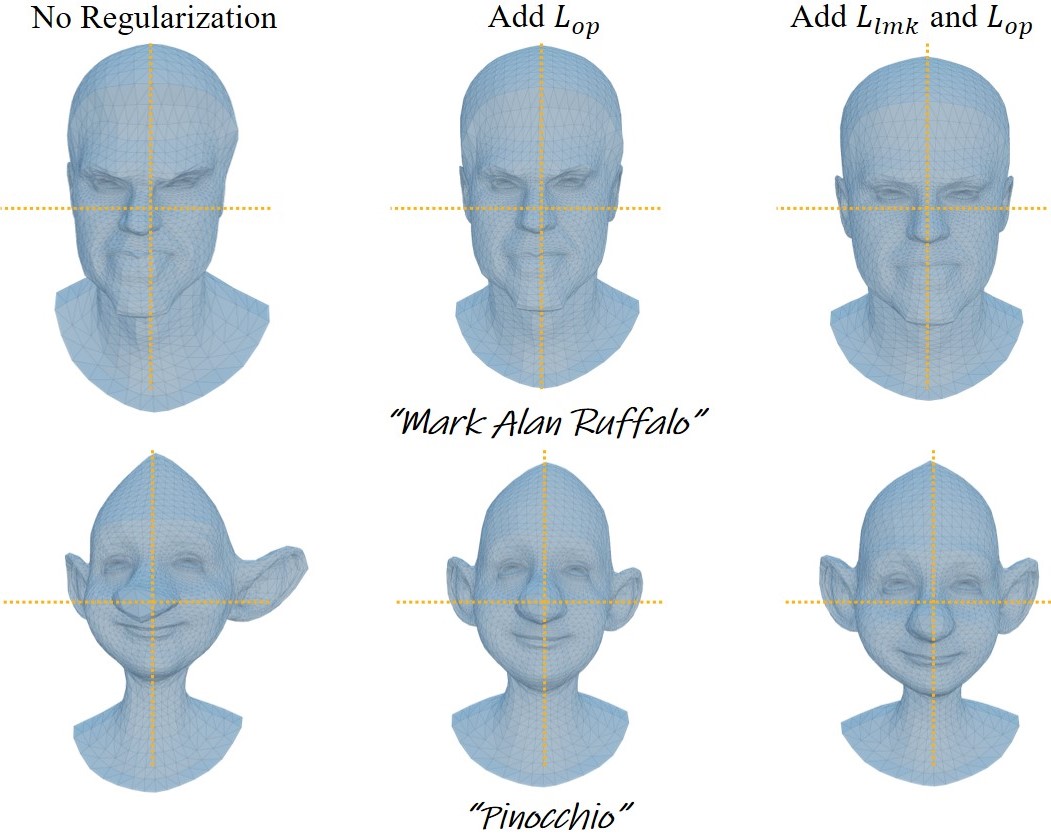}
\caption{Ablation study on regularization. The contour-based and landmark regularizations contribute to maintaining symmetric facial features and alignment with the pose of the input mesh.}
\label{Fig: Regularization Ablation}
\centering
\includegraphics[width=0.36\textwidth]{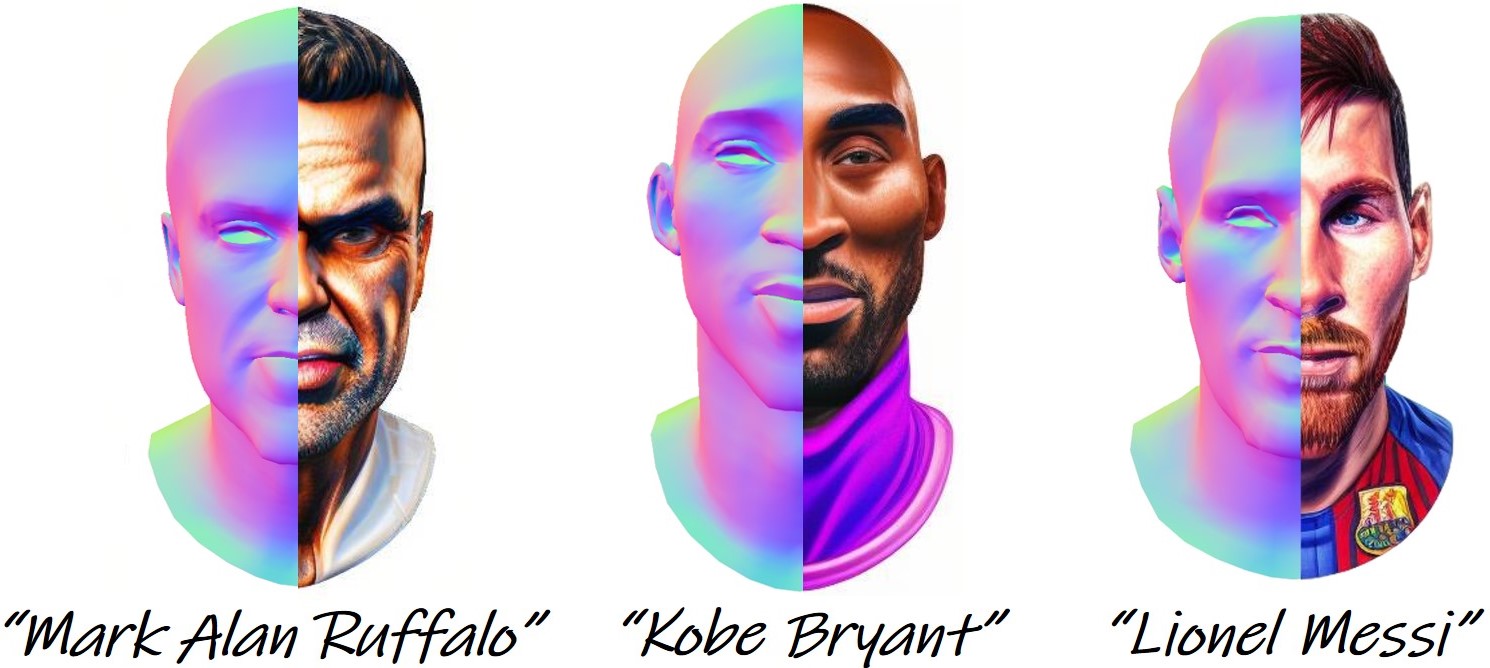}
\caption{Our deformation-based framework offers effective geometry consistency for further 3D applications.}
\label{Fig:Geometry Consistency}
\end{figure}

\textbf{Regularization.} 
In deformation optimization, facial features and head poses may deviate unexpectedly. Landmark regularization constrains generated meshes to the input shape's canonical space, and contour-based regularization contributes to reasonable head shapes. We demonstrate the effectiveness of our regularization designs in Figure~\ref{Fig: Regularization Ablation}.

\section{Applications}
\label{Sec:application}

\begin{figure}[!tbh]
\includegraphics[width=0.49\textwidth]{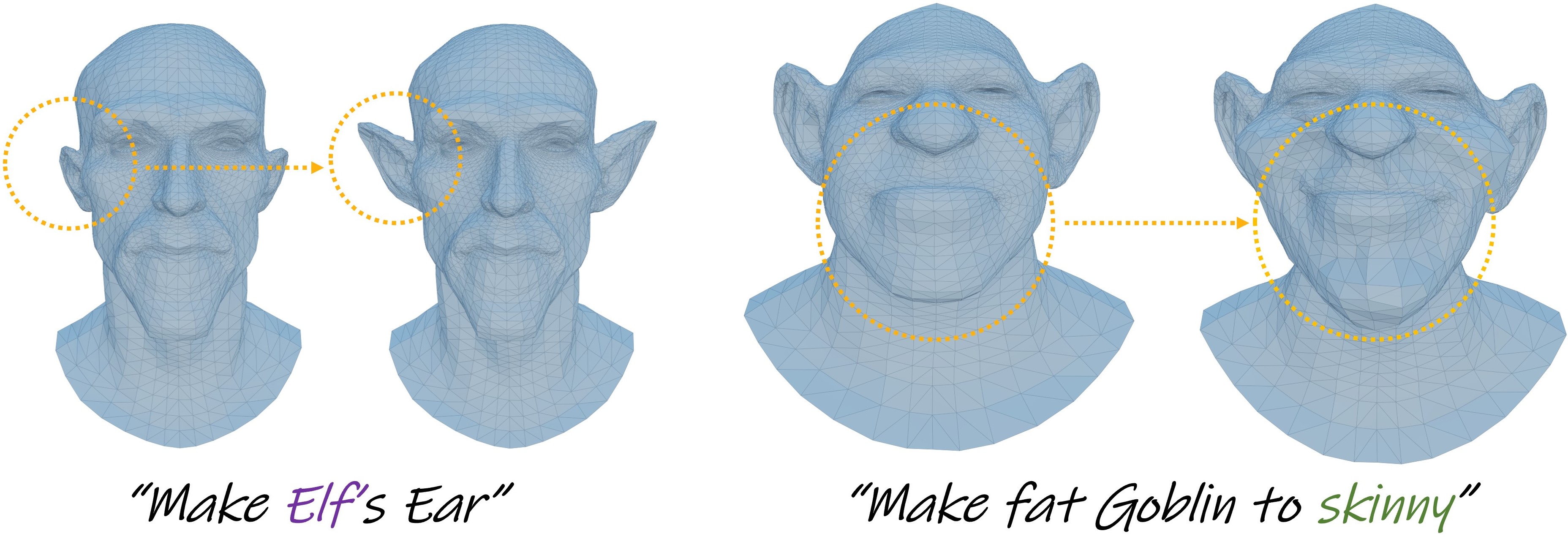}
\caption{Examples of local geometry editing. Our framework uses per-vertex segmentation information (Figure~\ref{Fig:semantics-coherence Demonstration}), enabling text-based mesh editing to alter avatars' local features.}
\label{Fig:Local Geometry Editing.}
\includegraphics[width=0.48\textwidth]{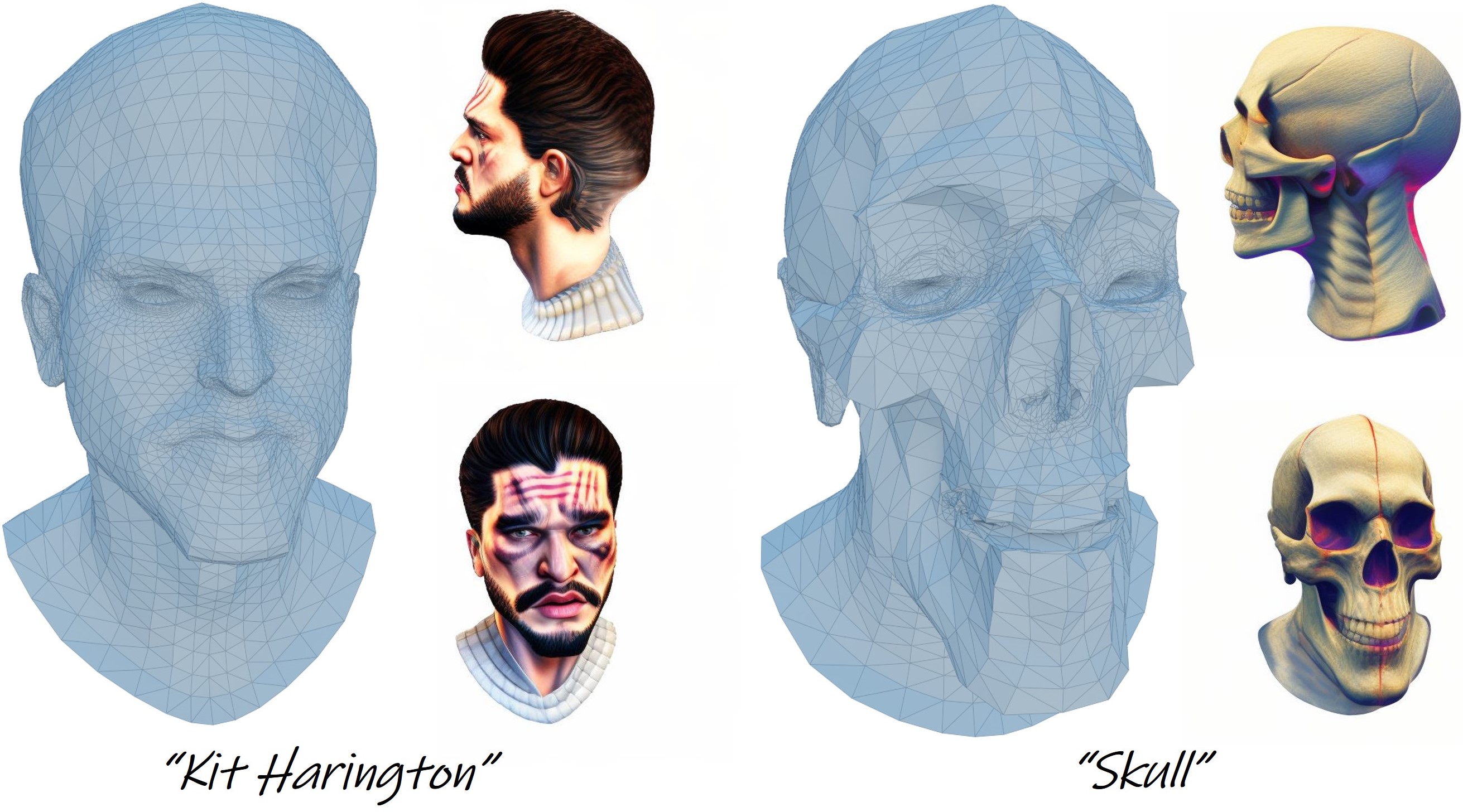}
\caption{More representative generation results.}
\label{fig:more_generations}
\includegraphics[width=0.48\textwidth]{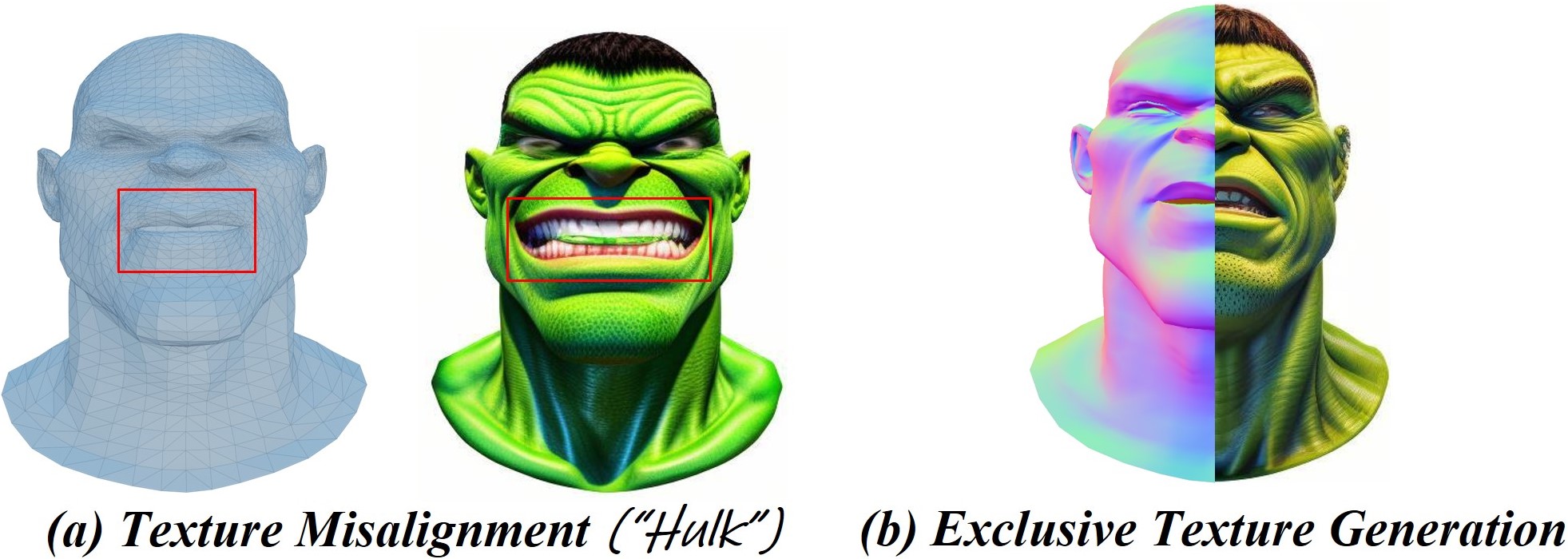}
\caption{A limitation of our method. For exaggerated figures like ``Hulk'', (a) though it does not impact further applications too much shown in Figure~\ref{fig:teaser}(c), perfect geometry-texture alignment is not guaranteed. (b) By fixing the geometry and solely applying texture optimizations, this artifact can be alleviated (e.g., TexGen~\cite{TexGen:ECCV:2024}).}
\label{fig:limitation_alignment}
\includegraphics[width=0.45\textwidth]{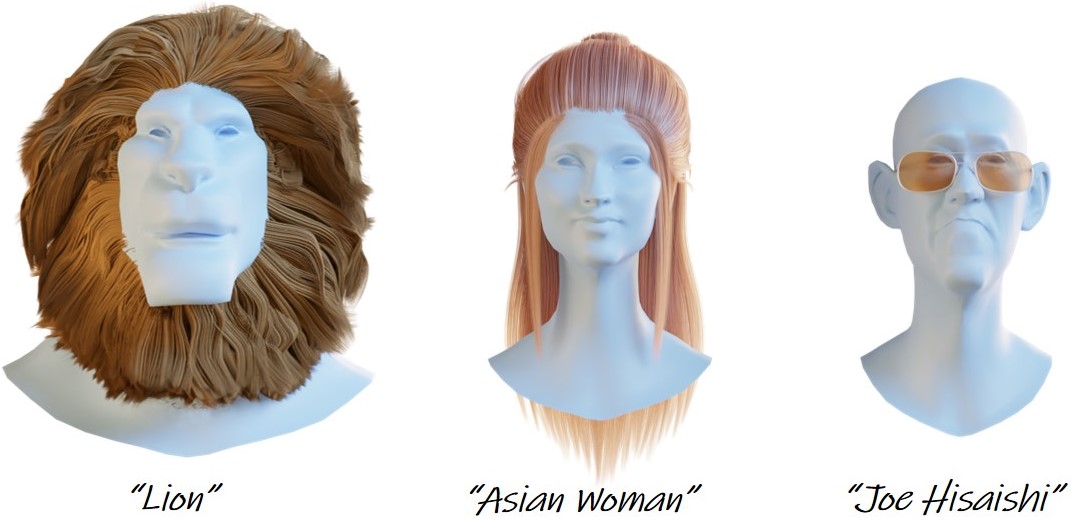}
\caption{Generated head avatars that can be manipulated (e.g., accessories) in graphics software without mesh post-processing.}
\label{fig:art_edit}   
\end{figure}

Our method produces head avatars that can be smoothly manipulated (e.g., animated via 3DMMs and text-based editing) and are compatible with graphics tools such as Blender and Maya. Compared with implicit representations~\cite{Fantasia3D:ICCV:2023, HeadSculpt:NIPS:2023}, our method enables artists to animate and edit head avatar assets for downstream applications (Figures~\ref{fig:teaser}(c) and~\ref{fig:art_edit}).

\textbf{Local geometry editing.}
Besides manipulating generated meshes in 3D software as artists usually do, our approach also supports efficient head editing with text descriptions (Figures~\ref{fig:teaser}(b) and~\ref{Fig:Local Geometry Editing.}). Only $H_i$ on certain parts is optimized given the text guidance. 

\begin{figure}[!tbh]
\includegraphics[width=0.48\textwidth]{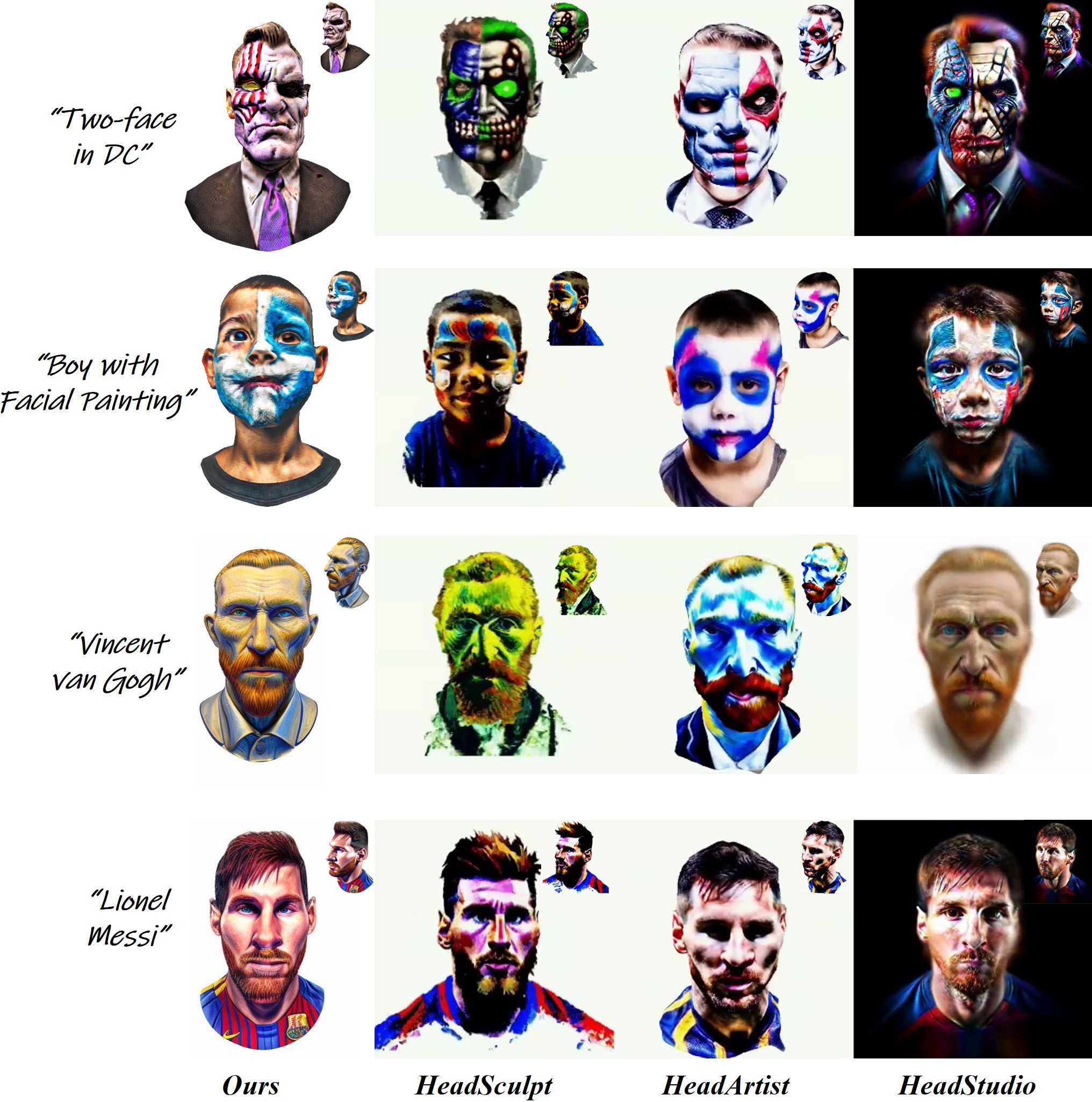}
\caption{More qualitative comparisons with recent methods that do not employ mesh representations for avatar generation.}
\label{fig:comp_implicit}
\includegraphics[width=0.485\textwidth]{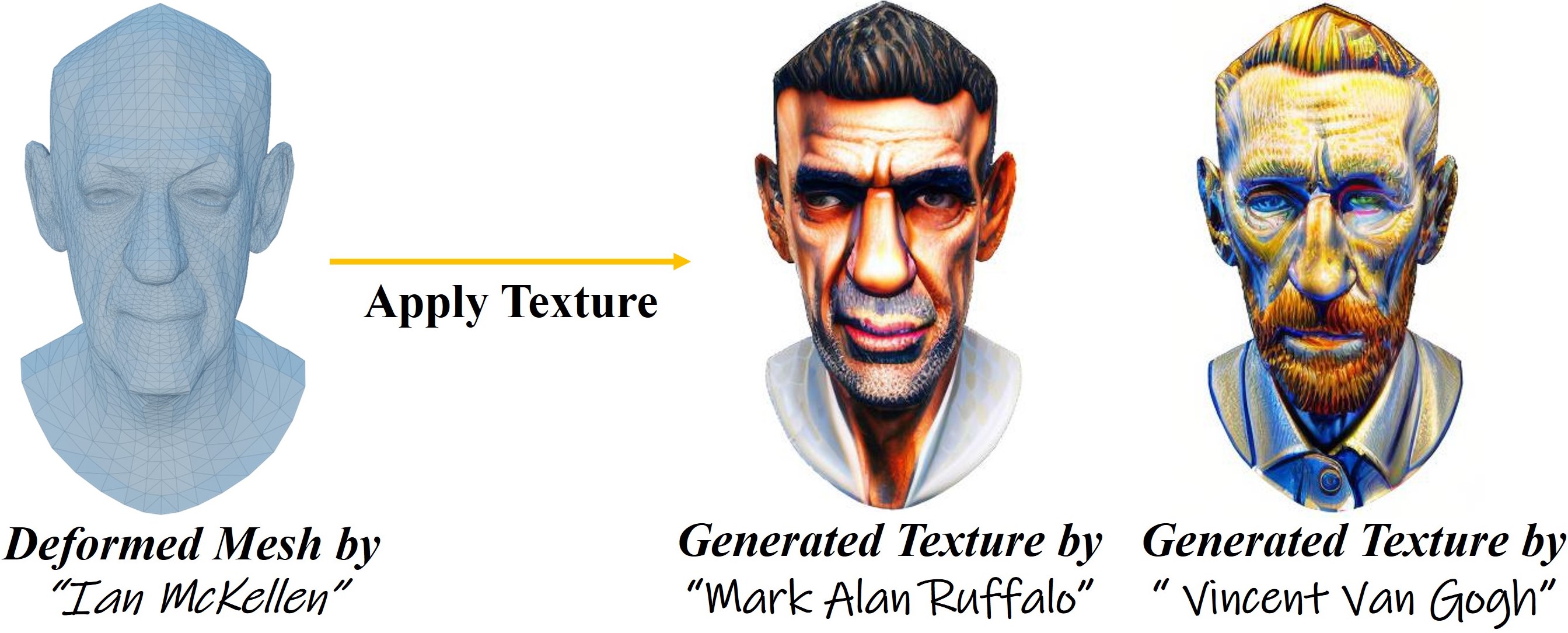}
\caption{Examples of texture transfer. Since the UV coordinates of the input template mesh are well preserved, various generated texture maps can be directly applied to the generated mesh.}
\label{Fig: extra transfer}
\includegraphics[width=0.48\textwidth]{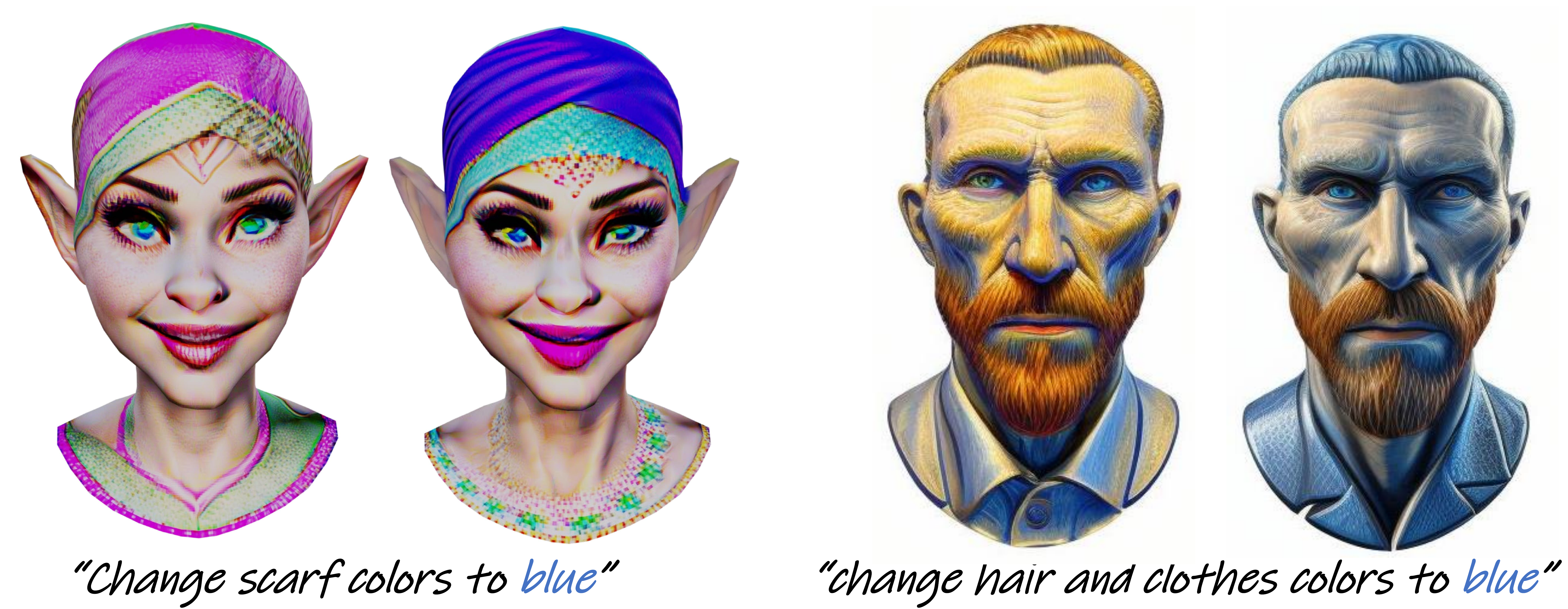}
\caption{Example of text-based texture editing. With extra text prompts, our method supports texture editing in partial regions.}
\label{Fig:Local Texture Editing.}
\end{figure}

\textbf{Texture editing through text guidance.}
Our approach encourages independent texture generation on a fixed geometry, which enables texture editing specified by the text input (Figure~\ref{Fig:Local Texture Editing.}). Some unexpected color artifacts may be introduced due to the noises from the diffusion model.

\begin{figure}[!tbh]
    \centering
    \includegraphics[width=0.48\textwidth]{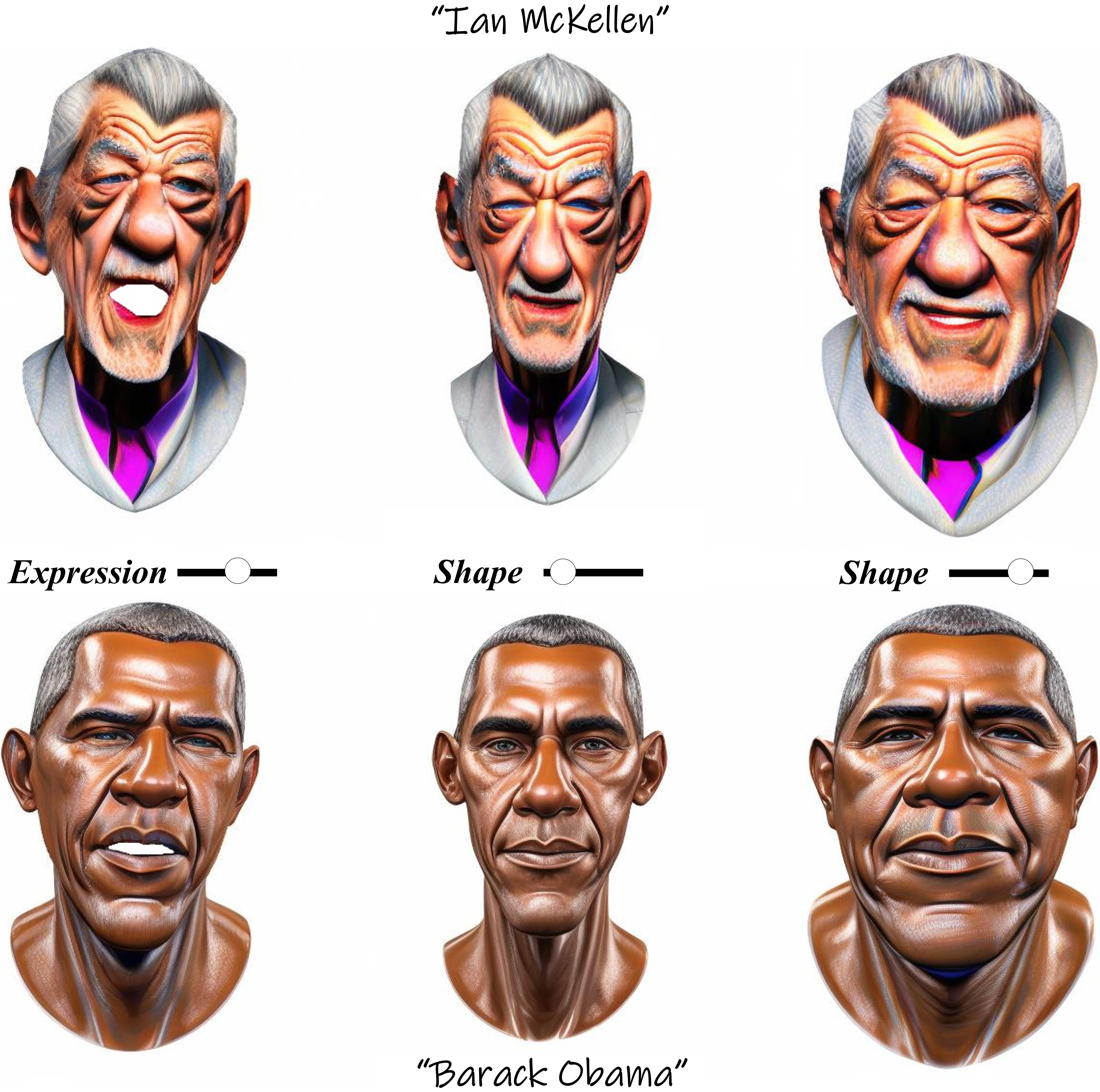}
    \caption{Example of attribute inheritance from FLAME~\cite{FLAME:TOG:2017}. If the input template mesh is from 3DMMs~\cite{3DMMs:SGP:1999}, users could manipulate the deformed mesh freely via morphable attributes such as expression and shape parameters.}
    \label{fig: more_3DMM}
\end{figure}

\textbf{Avatar animation.}
As shown in Figures~\ref{fig:teaser}(c) and~\ref{fig: more_3DMM}, we demonstrate that the rigging and blendshapes of the head mesh model after deformation are well-preserved for creating head animations and shape manipulations.

\textbf{Avatar generation from other template meshes.} As detailed in Section~\ref{Sec:method}, our method accommodates manifold meshes, not just specific models like FLAME. Figure~\ref{Fig:Fullbody Demonstration} exemplifies a "Superman" head creation without labeled 3D facial landmarks, using crafted 3D assets.

\section{Limitations and Future Work}
\label{Sec:futurework}

Our framework has a few limitations. Although many parametric head models do not contain eyeball mesh intrinsically (e.g., Facescape~\cite{facescape:CVPR:2020}, NPHM~\cite{nphm:CVPR:2023}, and BFM~\cite{BFM:FG:2018}), our experiments removed the eyeball mesh from FLAME and SMPL-X since Jacobians rely on the manifold mesh structure~\cite{NFR:SIGGRAPH:2023} (Figure~\ref{implementation}). A cage-based representation may offer deformations among multiple mesh instances simultaneously~\cite{NeuralCages:CVPR:2020}, enabling separately editable details like hair and accessories. Another minor limitation exists in that lighting isn't disentangled from diffuse colors, potentially causing noise and oversaturation artifacts. 

It is worth mentioning that the generated avatars exhibit relatively good geometry consistency with texture (Figure~\ref{Fig:Geometry Consistency}), but our method cannot guarantee an exact match between geometry and texture. For instance, eyeball appearance only shows in the texture, a problem that also occurs in TADA and HumanNorm. Stronger semantic labeling (e.g., canny edges~\cite{EASI-Tex:SIGGRAPH:2024}) may mitigate this issue. Another future focus is to learn head generation from a 3D sphere without any avatar shape priors~\cite{Largesteps:SIGGRAPH:2021}. 


\section{Conclusion}
\label{Sec:conclusion}
In this paper, we have presented a text-guided generative framework for crafting stylized 3D head avatars through learnable deformations. By employing Jacobians as an intermediate representation for vertex displacements and enhancing them with a learnable vector field, our method has improved the expressiveness of the deformed mesh. Consequently, learnable deformations are embedded into score distillation for geometry and appearance generation. Our regularization terms help retain 3D attributes from the source mesh to support seamless editing by artists. Experiments with various text prompts exhibit the remarkable performance of our framework. Our work can inform future research on effectively integrating mesh-based neural techniques into practical graphics pipelines, such as boosting the 2D and 3D deformation ability by applying our method to APAP~\cite{APAP:CVPR:2024}. 
\section{Acknowledgement}
This project is sponsored by CCF-Tencent Rhino-Bird Open Research Fund RAGR20230120.

{\small
\bibliographystyle{ieee_fullname}
\bibliography{reference}
}

\clearpage
\section*{Supplemental}
The supplementary material includes additional generation results and showcases applications such as text-guided geometry and texture editing.

\begin{figure}[!tbh]
\end{figure}

\end{document}